\documentclass[twocolumn,floatfix]{revtex4}
\usepackage{color}
\usepackage{epsfig}
\usepackage{graphicx}
\usepackage{amsmath}

\newcommand {\be}  {
\begin{equation}
}
\newcommand {\ee}  {
\end{equation}
}
\newcommand {\bea} {
\begin{eqnarray}
}
\newcommand {\eea} {
\end{eqnarray}
}

\begin{document}

\title{Strongly out-of-equilibrium columnar solidification during the Laser Powder-Bed Fusion additive manufacturing process}

\author{G. Boussinot$^1$, M. Apel$^1$, J. Zielinski$^2$, U. Hecht$^1$, J. H. Schleifenbaum$^{2,3}$}
\affiliation{$^1$Access e.V., Intzestr. 5, 52072 Aachen, Germany, \\
$^2$RWTH Aachen University - Digital Additive Production, Campus Boulevard, 52074 Aachen, Germany, \\
$^3$Fraunhofer Institute for Laser Technology ILT, Steinbachstr. 15, 52074, Aachen, Germany.}



\begin{abstract}

Laser-based additive manufacturing offers a promising route for  3D printing of metallic parts. We evidence experimentally a particular columnar solidification microstructure in a Laser Powder-Bed Fusion processed Inconel 718 nickel-based alloy, that we interpret  using phase-field simulations and classical dendritic growth theories.
Owing to the large temperature gradient and cooling rate, solidification takes places through dendritic arrays wherein the characteristic length scales, i.e tip radius, diffusion length and primary spacing, are of the same order. 
This leads to a weak mutual interaction between dendrite tips, and a drastic reduction of side-branching. The resulting irregular cellular-like solidification pattern then remains stable on time scales comparable to the complete melt pool solidification, as observed in the as-built material.

\end{abstract}

\maketitle

\section{Introduction}
Additive techniques where three-dimensional objects are produced layer by layer represent a promising route for the manufacturing of metallic alloys parts with complex geometries. Among these techniques, laser-based ones rely on the fusion of a certain amount of material, including the newly added powder, and its subsequent adhesion to the previously deposited material. The production of metallic parts via laser-based additive manufacturing processes is being implemented on industrial scales, however the understanding of the way to optimize mechanical properties is still poor. In particular, the best strategy for performing low-cost heat-treatments in order to homogenize the microstructure is still unknown. In this respect, the relation between process parameters, such as laser power or scanning velocity, and the solidification microstructure needs to be investigated.  

In the Laser-Powder-Bed-Fusion (LPBF) process, a laser beam typically 100 $\mu$m in diameter moves at large scanning velocities $V_s$, of the order of a meter per second. Its power is chosen high enough so that it completely melts the powder particles, whose size typically ranges around few tens of $\mu$m, as well as partly the previously deposited layer.
On the other hand, the power should not be chosen too high in order to avoid a porosity linked to the entrapment of vaporized melt (keyhole effect). Finite-Elements calculations \cite{macro} provide estimates of the cooling rates $\dot T$ within the solidification interval, of the order of 10$^6$ K/s. Moreover, due to the large laser beam scanning speed,  large temperature gradients $G$ reaching tens of millions of Kelvin per meter develop owing to the reduced heat diffusion length. 

The phase field model has proven its efficiency in reproducing the evolution of the solid/liquid interface during solidification processes.
It was originally developed for weakly out-of-equilibrium conditions. In this case, linear kinetic effects are present at the interface, with fluxes and driving forces being proportional through Onsager relations. Within the sharp-interface approach (transport equations in the bulk supplemented by boundary conditions at the interface), this proportionality is defined by the kinetic coefficients, and a formal link between the latter and the parameters of the phase field model may be found using the so-called thin-interface limit \cite{thin_int_karma, atc_karma, thin_int_moi}. 
 Here, however, at solidification growth velocities $V_g = \dot T/G \sim $ 10 cm/s, larger deviations from local equilibrium are expected at the interface. For this regime, the reduction of the phase field equations to a sharp-interface description is  lacking. Nevertheless, it was shown \cite{granasy, ghosh} that the phase field model provides a reasonable description of out-of-equilibrium phenomena such as solute trapping \cite{aziz}. 
 In this work, we aim at describing the solidification microstructure on the qualitative level, and we believe that the lack of correspondence between the phase field model and a sharp-interface description does not obliterate our conclusions. Moreover, we will show that characteristic microstructure features such as the inter-dendritic spacing obtained from simulations compare well with the experiments. 
  
In this article, we first present experimental observations of the solidification microstructure, pointing at inconsistencies with usual theories for solidification. We then present our phase field simulations that shed light on the origin of these inconsistencies, and we investigate the obtained solidification patterns in details.
 
\section{Experimental results} 
 
\begin{figure}[htbp]
\includegraphics[angle=0,width=150pt]{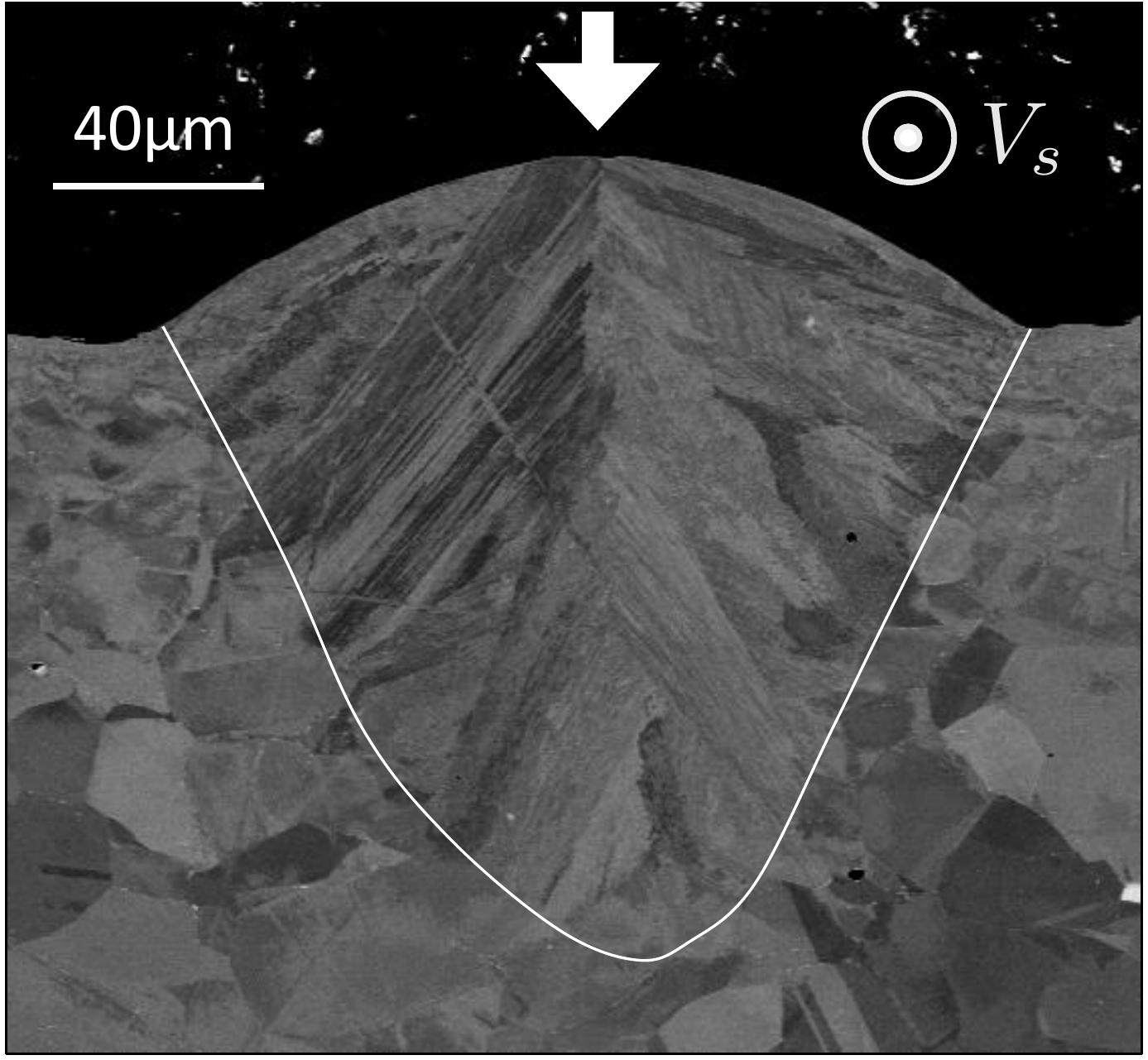}
\caption{\label{melt_pool} Longitudinal SEM image of a single-track LPBF experiment for the IN718 alloy (see text for further details). The material that underwent melting and solidification within the melt pool is enclosed by the white line.}
\end{figure}

\begin{figure}[htbp]
\includegraphics[angle=0,width=250pt]{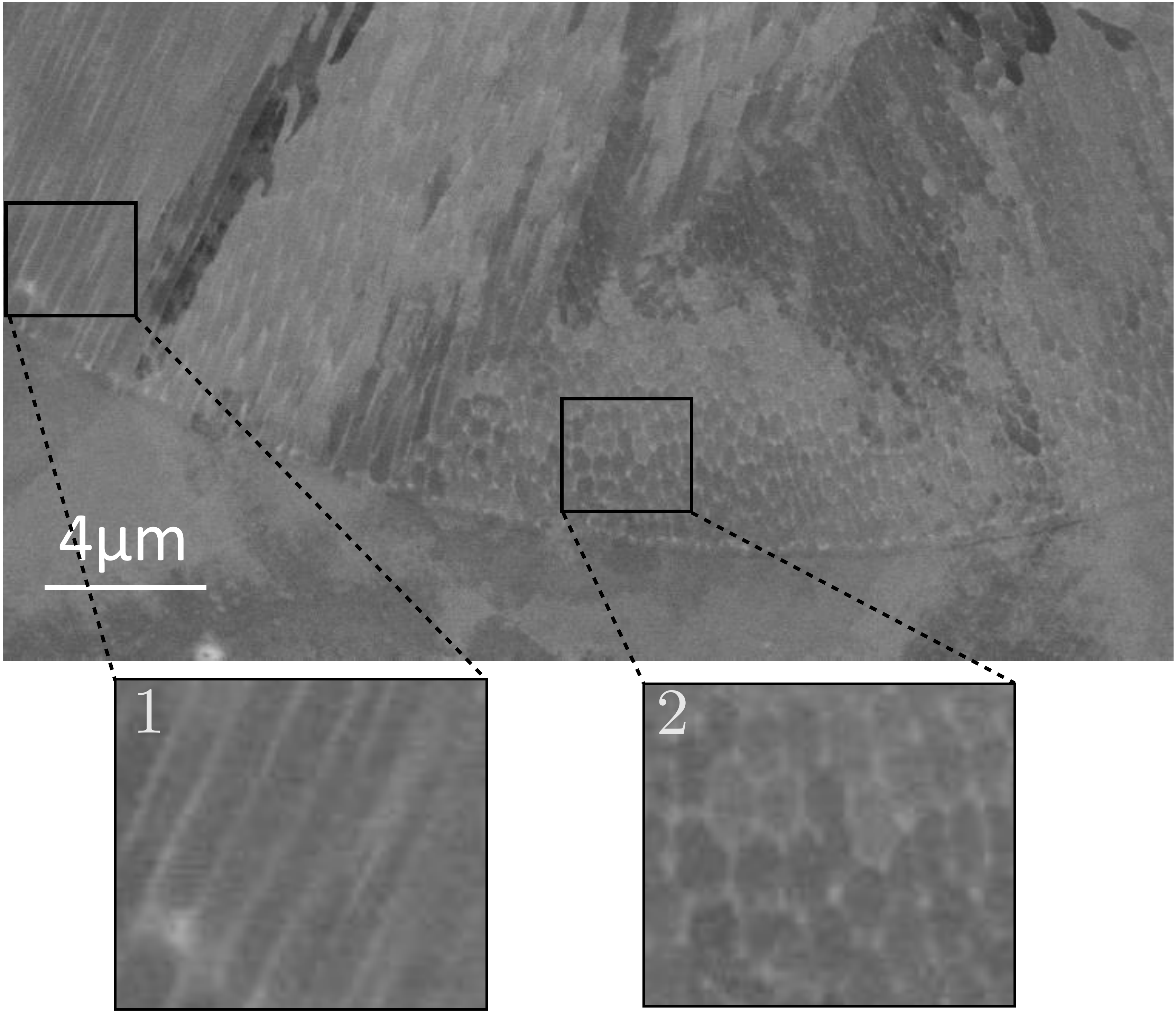}
\caption{\label{melt_pool_zoom} Zoom from Fig. \ref{melt_pool} at the bottom of the melt pool. The solidification stage leaves a white/light gray microsegregation pattern (absent in the unmolten material in the lower part of the figure). We enlarge two regions with a qualitatively different pattern.}
\end{figure}

 In order to study the solidification microstructure on the most fundamental level, we first performed a single-track experiment.
A 60 $\mu$m thick nickel-based IN718 superalloy powder bed is deposited on a forged IN718 base material, and a 285W laser beam traveling at a scanning velocity $V_s$ = 960 mm/s melts the powder and a certain amount of base material. 
 In Fig. \ref{melt_pool}, we present a SEM image from a section plane perpendicular to the laser track. The direction of the laser beam is denoted by the vertical arrow, and the scanning direction is perpendicular to the plane of the image. 
 The white line corresponds to the boundary of the melt pool and encloses the material that undergoes melting and solidification.
At several positions we see that the solidification is likely to take place without nucleation, but rather by epitaxy with grains from the unmolten substrate. 
At a smaller scale, we observe a white/light-grey microsegregation pattern. As can be seen in Fig. \ref{melt_pool_zoom}, it is clearly inherited from the solidification of the melt pool, since this microsegregation is completely absent in the region corresponding to the unmolten base material  in the lower part of the picture. 
We enlarge two regions with a qualitatively different microsegregation pattern. In region 1 in the left panel, the microsegregation pattern is unidirectional, indicating that the angle between the growth direction and the plane of the image is almost zero. In region 2, the pattern consists of a rather isotropic two-dimensional network indicating a growth direction making a significant angle  with the plane of the image. No side branching is observed (although it is observed at few instances when analyzing a larger area), pointing to a cellular growth structure. The characteristic length scale associated with this cellular structure, i.e. the periodicity, is  between 300 and 600 nanometers. 

The fact that,  in the two enlarged regions, the growth is cellular with significantly different growth directions challenges the usual theories of solidification. Indeed, being at the bottom of the melt pool implies that the temperature gradient lies mostly within the plane perpendicular to the laser track, i.e. within the plane of Fig. \ref{melt_pool_zoom}. Thus the growth direction in region 1 is well-aligned with the temperature gradient, while the growth direction in region 2 is significantly misaligned. 
However, cellular growth is known to take place under large $G/V_g$ ratios with the growth direction aligned with the temperature gradient. Therefore, the growth direction of the cellular structure in region 2 shows a clear inconsistency with previous observations \cite{akamatsu, pocheau} and theoretical results \cite{levine}.
In addition, we identify grains in Fig. \ref{melt_pool} that span the whole solidified area, starting at the melt pool boundary (white line) and reaching  by a straight line the axis of symmetry of the melt pool. This suggests that the projection of the growth direction on the image plane does not change during the whole solidification of the melt pool for these grains, while the direction of the temperature gradient of course varies significantly on this scale. This contradicts again with a cellular growth direction following the temperature gradient. 

\begin{figure}[htbp]
\includegraphics[angle=0,width=220pt]{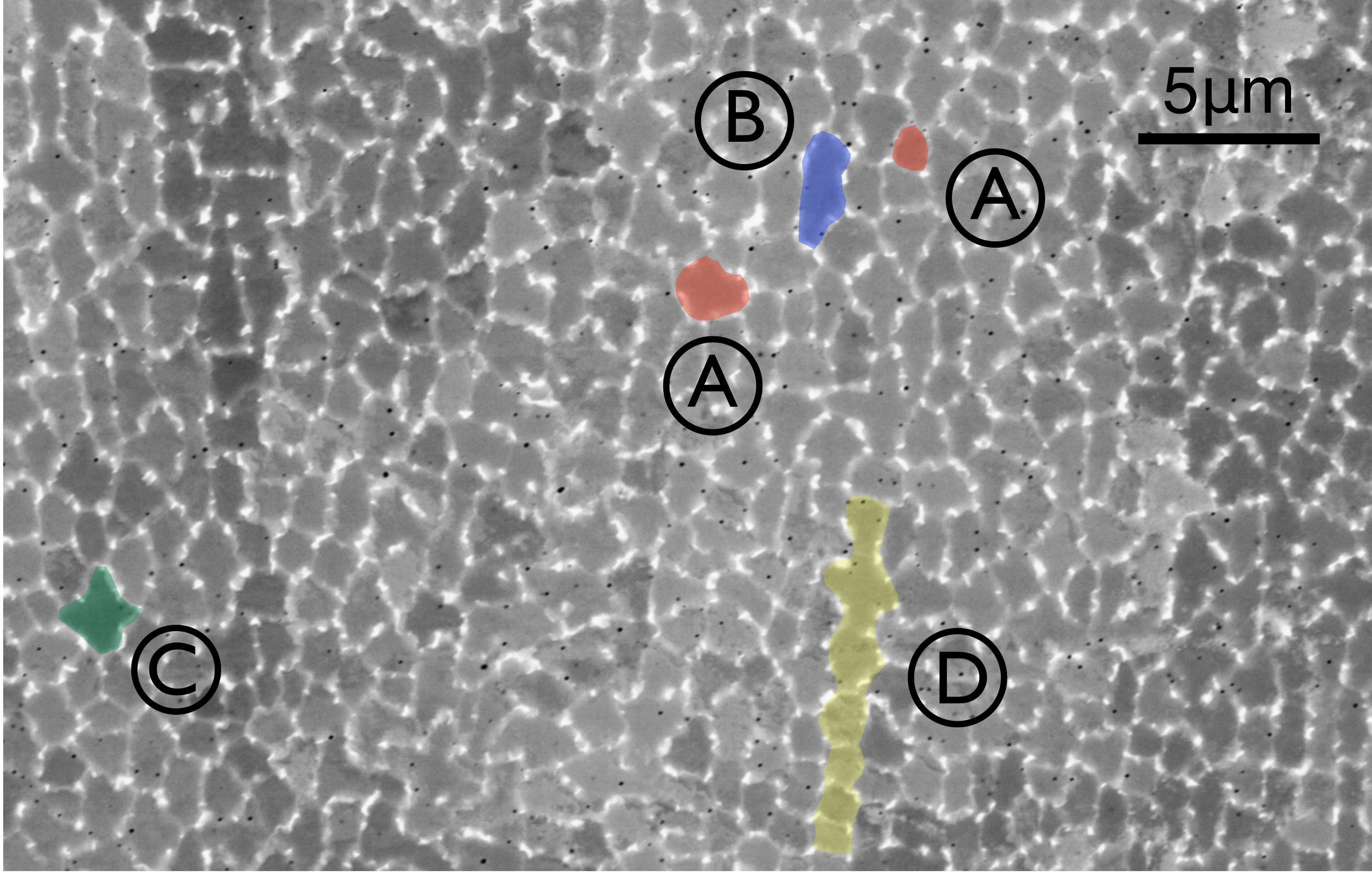}
\caption{\label{transversal} Transversal SEM image of a LPBF-processed IN718 alloy showing the solidification pattern.}
\end{figure}

Let us now focus on a transversal cut normal to the build direction of a LPBF-processed IN718 part, i.e. a cut normal to the direction of the laser beam. The power and the scanning velocity are the same as for the single track experiment, and we present in Fig. \ref{transversal} a corresponding SEM micrograph. In white/light gray color we distinguish the inter-cellular regions.
Here, the typical inter-cell distance is larger than in Fig. \ref{melt_pool_zoom}, i.e. around 1-2 $\mu$m, meaning that the thermal conditions are different (we are probably not at the bottom of a melt pool like in Fig. \ref{melt_pool_zoom}). Noticeably, the shape and width of the cells and their arrangement are irregular. We highlight in red two cells with a rather similar compact shape but with a linear size differing by a factor about two (A). In blue, we highlight a cell whose linear size in one of the directions is about three times larger than in the other direction (B), and in green, we highlight a cell with a shape that is typical of dendritic growth in nickel-based alloys (C). Finally, as mentioned just above, while the cellular network is quite irregular, i.e. deviates strongly from a periodic array, we highlight in yellow few cells that nevertheless show some alignment (D).

The first idea to explain these irregularities is that the solid/liquid interface is experiencing time dependent thermal conditions, with the cellular structure thus continuously changing. As an example we may observe cells that exhibit a very small cross section compared to others due to the so-called termination, i.e. when they are left behind the growth front and subsequently undergo ripening within the mushy zone.
However, we performed phase field simulations of the solidification of the IN718 alloy, and our simulations tend to show that the irregularity of the microstructure that is observed experimentally may not be necessarily due to transient thermal conditions.

\section{Phase field simulations}
\begin{figure}[htbp]
\includegraphics[angle=0,width=200pt]{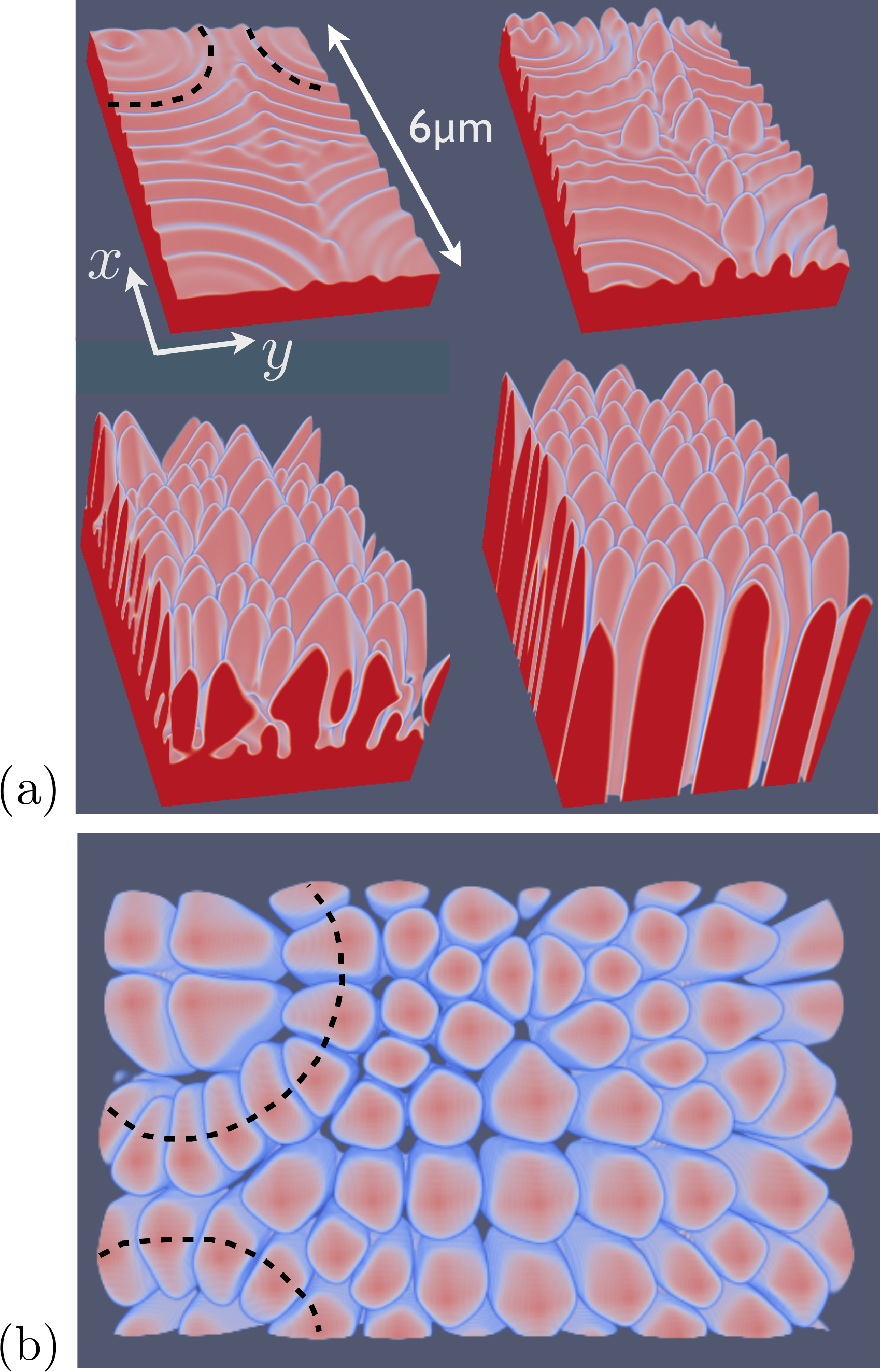}
\caption{\label{3d} Phase field simulation of the cellular growth in IN718 at $V_g$ = 4 cm/s and $G = 10^5$ K/cm: a) time evolution for $t=0.05, 0.075, 0.1, 0.15$ ms; b) top view (thermal gradient perpendicular to the plane of the image) of the solidification pattern, stable for long times, in the order of complete solidification of the melt pool. The cellular arrangement shows some irregularity and some correlation with the initial conditions, as illustrated by the dashed lines.}
\end{figure}

The phase field simulations are performed using the MICRESS software \cite{micress}, based on a multi-component phase field model \cite{janin} that is coupled to the TCNI8 Thermo-Calc \cite{thermocalc} database. The composition of the alloy was chosen as follows, in wt\%: 17.64 Fe, 19.00 Cr, 5.13 Nb, 3.05 Mo with Ni as balance. 
In Fig. \ref{3d}, we present a three-dimensional simulation where the growth velocity is $V_g = $ 4 cm/s and $G = 10^5$ K/cm.  
The thermal gradient $G$ is considered homogeneous in space and constant in time (so-called frozen temperature approximation), and diffusion is neglected in the solid phase. 
The crystalline axis of the solid make ($\varphi = 5^\circ,\theta = 5^\circ, \psi = 5^\circ$) Euler angles with respect to the thermal gradient. 
 The discretization grid spacing is 20 nm and the simulation is initialized with an almost flat solid/liquid interface.
In (a), we present a side-view of the microstructure at $t=0.05, 0.075, 0.1, 0.15$ ms, with the direction of the thermal gradient being perpendicular to the initial flat front.
At 0.05 ms, the flat front destabilizes and the symmetry of the initial conditions becomes visible. At 0.075 ms, few well-developed cellular protrusions have  appeared. At 0.1 ms, all the simulation domain is invaded by cells, although all the tips are not at the same temperature. At 0.15 ms, the cells that were in the rear of the leading ones at 0.1 ms have reached the latter. This cellular arrangement then remains mostly unchanged for the rest of the simulation, i.e. 1.4ms (corresponding to 56 $\mu$m of growth, which is of the order of the size of the melt pool). 
A drift of the whole structure is however observed due to the rotation of the crystalline axis with respect to the thermal gradient. If the growth direction strictly follows  the cubic axis of the crystal, the drift velocity in the $x$  direction is $V^0_x = V_g \sin \varphi \sin \theta$ and the drift velocity in the $y$  direction ($z$ direction is aligned with the thermal gradient) is $V^0_y = -V_g \cos \varphi \sin \theta$. We find in our simulation a drift velocity in the $x$ direction $V_x \approx 0.88 \;V^0_x$ and a drift velocity in the $y$ direction $V_y \approx 0.90\; V^0_y$, which means that the growth direction is not strictly following the cubic axis of the crystal but is close to it. 
In Fig. \ref{3d}b, we present a top view (thermal gradient perpendicular to the plane of the image) of the cellular structure in this quasi-steady state. We see that the shape, size and organization of the cells show irregularity, as observed in the experiment. We note also the circular arrangement illustrated by the dashed line, and inherited from the symmetry of the initial conditions (see the dashed lines in Fig. \ref{3d}a).

From the number of cells in the simulation, we derive an average cell spacing $\lambda$. The observed  density is 2.6 cells/$\mu$m$^2$, i.e. $\lambda = 1/\sqrt{2.6} \; \mu$m $\approx$ 620 nm. Thus $\lambda$ falls in the range observed experimentally, i.e. from 300nm to 2$\mu$m. This is quite satisfactory in view of the absence, mentioned in the introduction, of a quantitative phase field model (reducible to a sharp-interface model) for such velocities and of the uncertainty on material parameters such as the diffusion coefficient in the liquid $D$ (here we chose $D = 2 \times 10^{-9}$ m$^2$/s) or the interface energy (we chose 0.245 J/m$^2$ and an anisotropy of 2\%).
 However, although one may define an average spacing $\lambda$, we clearly see in Fig. \ref{3d} that $\lambda$ does not provide sufficient informations to characterize the solidification microstructure, where we find cells with very different shapes and widths. 
 Another important observation is that the cells do not show secondary arm branching, as observed in the experiments. Let us now discuss and interpret these features of the solidification microstructure.

\section{Discussion}
Within the classical dendritic growth theory, the P\'eclet number $P_\rho=\rho/l_D$ is related to the deviation from equilibrium $\Delta = (T_L-T_0)/(T_L-T_S)$ through Ivantsov relation \cite{ivantsov}. Here $\rho$ is the tip radius of curvature (rigorously the radius of the Ivantsov parabola) and $l_D = 2D/V_g$ is the diffusion length. $T_L$ is the liquidus temperature, here 1630K for our 5 elements IN718 alloy, $T_S$ is the solidus temperature, here 1570K, and $T_0$ is the tip temperature.
In three dimensions, the Ivantsov relation reads $\Delta = P_\rho \exp(P_\rho) \int_{P_\rho}^\infty x^{-1} \exp(-x) dx$, while in two dimensions it reads $\Delta = \sqrt{\pi P_\rho} \exp(P_\rho) \text{erfc}(\sqrt{P_\rho})$. Then, the selection theory, that involves the anisotropy of interface energy, provides the additional condition to find $\rho$ (see Ref. \cite{book_hmk} and references therein). 
In the weakly out-of-equilibrium (WOE) regime, we have  $\Delta \ll 1$ and $P_\rho \ll 1$.
For the simulated LPBF thermal conditions, the tip radius of curvature, displayed in a pronounced red color in Fig. \ref{3d}, is of the order of the diffusion length $l_D = 100$ nm, and thus $P_\rho \sim 1$. In this case, the Ivantsov relation implies that $\Delta$ and $1-\Delta$ are of order unity, and we refer to this regime as the strongly out-of-equilibrium (SOE) regime. When $0< 1-\Delta \ll 1$, the solidification structure is a nearly flat front with $P_\rho \gg 1$ \cite{langer, efim} that approaches the regime of so-called 'rapid solidification' with $\Delta > 1$. 
For the simulation in Fig. \ref{3d}, we find $\Delta \approx 0.5$, that corresponds to $P_\rho \approx 0.6$ according to the 3D Ivantsov relation. 

It is well-established that the morphology of the growth front in directional solidification under a thermal gradient $G$ is mostly related to the dimensionless number $\beta = l_D/l_T$ where $l_T$ is the solidification length, i.e. $l_T = (T_L-T_S)/G$.
When $\beta$ is larger than some critical $\beta_c$ of order unity, i.e. $\beta/\beta_c > 1$, the Mullins-Sekerka instability is inhibited \cite{langer_RMP} and the flat front solution is linearly stable. When $\beta/\beta_c \ll 1$, i.e. in the isothermal regime, the solidification is equiaxial dendritic. When $\beta/\beta_c \lesssim 1$, the growth is columnar. 
Steady columnar growth is hardly achieved owing to the existence of a continuous family of stable periodic steady-state solutions, existing within a finite interval of spacing/periodicity. Thus in a spatially extended system, long wave-length variations of average spacing as well as local variations around the average value are present. 
At the lower limit of the stable spacing interval, the above-mentioned termination (also called elimination) is expected to be the mechanism by which an array, too dense to be stable, increases its spacing in order to gain stability. 
At the upper limit, the tertiary branching mechanism \cite{tourret, karma_new}, by which a tertiary dendrite side arm develops into a primary one, is expected to allow for a decrease of the spacing. 
On the other hand, for an organization of the array without change in the average spacing, the phase diffusion process \cite{ph_diff_rappel, ph_diff_karma} operates. 

In the following we want to show that the solidification microstructure that is obtained, experimentally and from simulation, for the LPBF process is related to the particular growth conditions denoted as the SOE regime above. 
We consider the two P\'eclet numbers $P_\rho$ and $P_\lambda = \lambda/l_D$. We define $\chi = P_\rho/P_\lambda = \rho/\lambda$ that may not be, for geometrical reasons, much larger than unity. We assume that side branching operates for $\chi \ll 1$, while for $\chi \sim 1$, the growth structure is cellular. This distinction is supported by the admitted theory for noise-induced side-branching \cite{efim_side_branch} on a non-axisymetric 3D dendrite \cite{efim_3d} for which secondary side branches develop at distances from the tips of order $\rho$. 
Thus when $\rho \ll \lambda$, secondary branches have the ability to grow freely possibly allowing for higher-order branching. In opposition, when $\rho \sim \lambda$, primordial secondary branches from neighboring primary trunks impinge, thus inhibiting their further development and higher-order branching.

In the WOE regime where $P_\rho \ll 1$, $\chi \ll 1$ corresponds to the classically observed columnar dendritic growth. The stability of the dendrite tips is linked to the anisotropy of the solid/liquid interface energy (while the tips are unstable against splitting in absence of anisotropy \cite{akamatsu_tip_split}), and the growth direction, depending on $P_\lambda$, tends to be aligned with the crystalline cubic axis, i.e. the direction of minimum interface stiffness. The diffusion fields of the dendrite tips  overlap weakly and the spacing between dendrites is of the order of the diffusion length $l_D$, i.e. $P_\lambda \sim 1$. 
Owing to $\chi \ll 1$, side branches develop and a decrease of the dendrite spacing is thus possible through the tertiary branching mechanism. 
When $\chi \sim 1$, the growth corresponds to the cellular structure that operates for $1 - \beta/\beta_c \ll 1$ and arises as a bifurcation from the flat front solution. 
Owing to $P_\lambda \ll 1$, the diffusion fields ahead of the tips strongly overlap and the phase diffusion process is especially effective, tending to organize the array through an homogenization of the cellular spacing. 
As mentioned when describing the experimental results, the growth direction of the cells is then aligned with the thermal gradient and the anisotropy of interface energy does not play a significant role. 
In opposition, when $P_\lambda \sim 1$ as for a dendritic array, the phase diffusion coefficient is significantly smaller since the Green's function for the diffusional interaction between two points separated by a distance $x$ is proportional to a decaying exponential, with its argument being $x/l_D$. Then the anisotropy of interface energy may influence the growth direction of the dendrites.

\begin{figure}[htbp]
\includegraphics[angle=0,width=230pt]{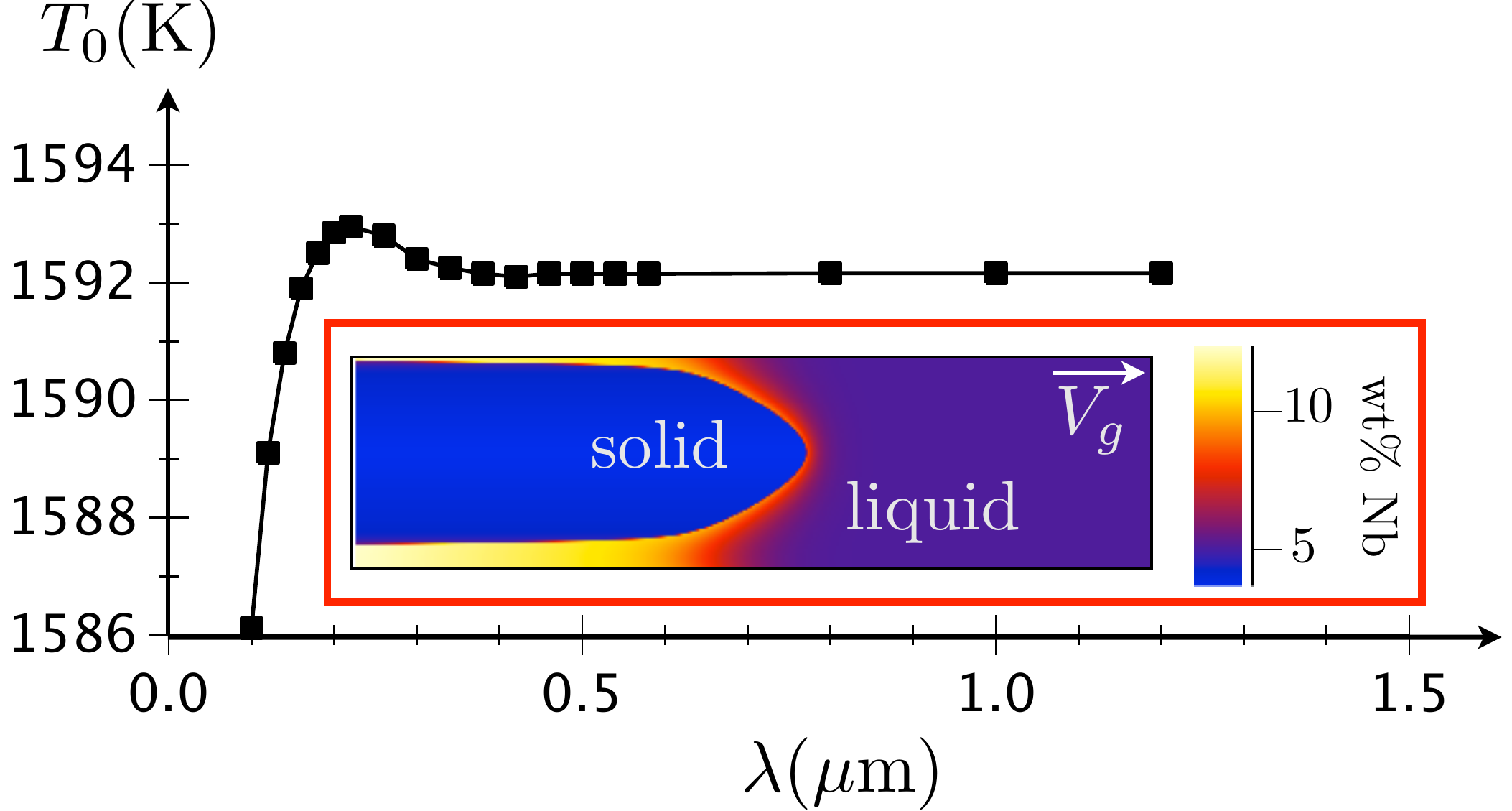}
\caption{\label{undercooling_vs_lambda} Single cell in a channel: temperature of the tip $T_0$ as a function of the channel width $\lambda$ for $V_g = 4$ cm/s and $G=2 \times 10^5$ K/cm. Here $l_D = 0.1 \mu$m.}
\end{figure}

Let us now analyze the SOE regime for which $P_\rho \sim 1$. A particularity of this regime concerns the fact that $\chi$ may not be much larger than unity, i.e. $\lambda$ should be larger than $\rho$. When $P_\rho$ increases, the minimum value of $P_\lambda$ thus increases in the same way. 
 As a consequence, the interval of stable spacing in the SOE regime is shifted to larger $P_\lambda$ compared to the WOE regime. We performed two-dimensional simulations of a single cell in a channel in the SOE regime, with $V_g = 4$ cm/s, $G=2 \times 10^5$ K/cm, the cubic axis of the crystal aligned with the thermal gradient, and otherwise the same physical parameters as for Fig. \ref{3d}. The grid spacing is set to 2 nm. We present in Fig. \ref{undercooling_vs_lambda} the tip temperature $T_0$ as a function of the channel width $\lambda$ in the stable range together with a snapshot of the Nb concentration field for the case $\lambda = 800$ nm. We see that $P_\lambda$ ranges from 1 to 12 in the stable regime ($l_D=$ 0.1 $\mu$m).  
In comparison,  in Ref. \cite{karma_new}, stable $P_\lambda$ in a succinonitrile-d-camphor alloy for a 15$^\circ$ misoriented crystal was found (pay attention to the different definition of $P_\lambda$ in \cite{karma_new}) to range from 0.17 to 2.4 at $V_g$ = 4 $\mu$m/s and from 0.38 to 7.1 for a five times larger velocity (see Table II in \cite{karma_new}). 
At the upper limit of the stability range where $P_\lambda = 12$ in our simulations, we find  $\chi = \rho/\lambda \approx 0.056$ with $\rho \approx 67$ nm being evaluated using the phase field contour at the tip. 
Although $\chi$ is rather small, the limit $\chi \ll 1$ implies $\lambda \gg l_D$ when $P_\rho \sim 1$, which is not a stable situation since the diffusion fields around the tips do not overlap. Thus,  on the theoretical level, $\chi$ is of order unity in the SOE regime for columnar solidification.
In practice, whether using a $\chi \ll 1$ situation or a flat front as initial condition, the system rapidly evolves towards a $\chi \sim 1$ regime, with typically $P_\rho \approx 0.6$ and $P_\lambda \approx 6$ as for our 3D calculation in Fig. \ref{3d}, yielding $\chi \approx 0.1$.
 This explains why we do not observe pronounced side branching in the experimental and simulated microstructures.
In comparison, $\chi \approx 0.01$ for a stable array of dendrites  growing at $V_g$ = 30 $\mu$m/s with well-developed secondary arms in a Al-Cu4wt\% alloy \cite{nous}. 
 However, side branching may still occur during SOE columnar growth when, locally, the large spacing limit of stability is overcome, for example at a divergent grain-boundary as we will see below.  
As a side remark, it is also probably worth mentioning that, using the same thermal gradient and almost flat initial conditions as in Fig. \ref{3d}, we find that the classical law describing diffusion-controlled processes $\lambda^2 V_g = C^{\text{ste}}$ holds in the SOE regime when we change $V_g$ within one order of magnitude. 

\begin{figure}[htbp]
\includegraphics[angle=0,width=\linewidth]{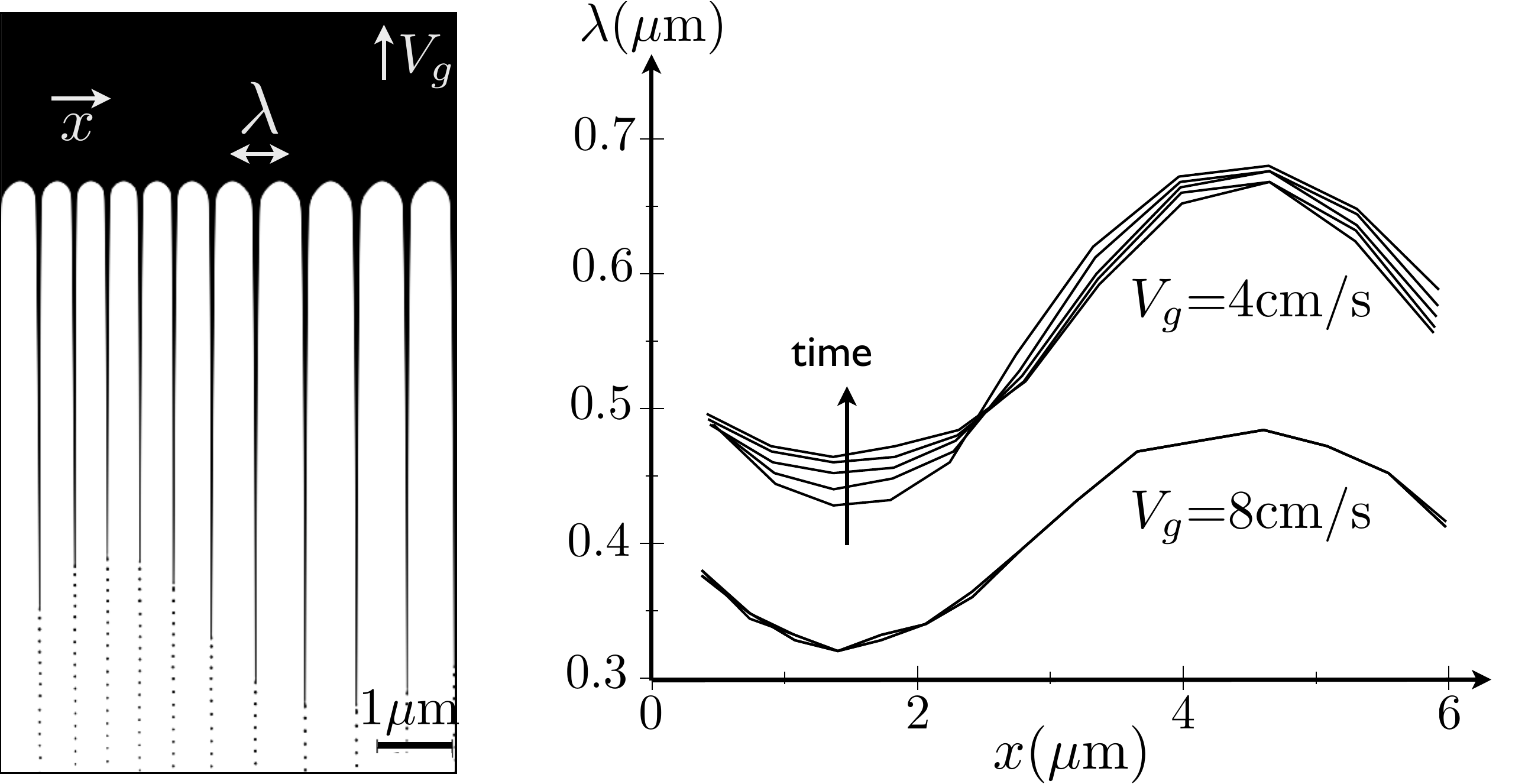}
\caption{\label{phase_diff} Left: cellular array exhibiting a sinusoidal modulation of the spacing $\lambda(x)$. The amplitude of the modulation decreases in time according to the phase diffusion equation $\dot \lambda = D_{PD} \partial^2 \lambda/\partial x^2$. Right: $\lambda(x)$ at times $t=$ 0.2, 0.3, 0.4, 0.5, 0.6 ms for $V_g=4$ cm/s, and at times $t=$ 0.125, 0.175, 0.225, 0.275 ms for $V_g=8$ cm/s. $D_{PD}$ is much larger in the first case than in the second.}
\end{figure}

Let us now focus on phase diffusion. Due to the larger values of stable $P_\lambda$ compared to the WOE regime, the phase diffusion process is largely suppressed in the SOE regime. This is illustrated by the fact that the undercooling of the tip becomes constant for $P_\lambda \gtrsim 4$ in Fig. \ref{undercooling_vs_lambda}. This indicates that the state of the tip is very weakly depending on the size of the channel, and therefore that the tips in the corresponding ideal periodic array are very weakly interacting. 
We performed additional two-dimensional simulations with $G=2 \times 10^5$ K/cm for $V_g = 4$ cm/s and $V_g = 8$ cm/s in order to estimate the influence of the growth velocity on the phase diffusion coefficient. In the spirit of the Appendix in Ref. \cite{karma_new}, we looked for the evolution of a cellular array presenting a sinusoidally modulated spacing, as shown in the left panel of Fig. \ref{phase_diff}. Here, the anisotropy of interface energy is set to 0.3\% and the cells' tip is more round than in Fig. \ref{undercooling_vs_lambda}.
According to the phase diffusion equation $\dot \lambda = D_{PD} \partial^2 \lambda/\partial x^2$ where $D_{PD}$ is the phase diffusion coefficient, the spacing tends to homogenize with a decrease in the amplitude of the spacing modulation. 
The lateral size in the $x$ direction of the simulation box is 6$\mu$m for both growth velocities, and we impose 11 cells in the case $V_g=4$ cm/s (Fig. \ref{phase_diff}) and 15 cells  in the case $V_g=8$ cm/s (not shown).
The corresponding average spacing is chosen according to the results of simulations with the same parameters, but with an almost flat initial condition. These simulations exhibit the $\lambda^2 V_g = C^{\text{ste}}$ law and the average spacing thus differs (only roughly due to finite size effects) by a factor $\sqrt{2}$ between $V_g=4$ cm/s and $V_g=8$ cm/s. 
In the right panel of Fig. \ref{phase_diff}, we plot the evolution of the spacing distribution. To each mid-point between two tips, we assign the corresponding distance between them and periodic boundary conditions are imposed laterally. 
For $V_g = 4$ cm/s, we present $\lambda(x)$ for times $t=0.2, 0.3, 0.4, 0.5, 0.6$ ms. For $V_g = 8$ cm/s, we present $\lambda(x)$ for times $t=$ 0.125, 0.175, 0.225, 0.275 ms. We see that, while the damping of the spacing modulation is substantial for $V_g=4$ cm/s, the spacing evolution is almost frozen for $V_g=8$ cm/s. We find $D_{PD} \approx$ 300 $\mu$m$^2$/s for $V_g=4$ cm/s, while for $V_g = 8$ cm/s, $D_{PD}$ is at least one order of magnitude smaller (the phase diffusion coefficient is in this case difficult to assess due to the small time variations of $\lambda$). The difference in (average) $P_\lambda$ is approximately 2.5 ($P_\lambda \approx 5.5$ for $V_g=4$ cm/s and $P_\lambda \approx 8$ for $V_g=8$ cm/s). Thus a dependence of $D_{PD}$ on $P_\lambda$ such as $D_{PD} \propto \exp(-P_\lambda)$ seems to be plausible since $\exp(-2.5) \approx 0.08$. 

This study of the phase diffusion coefficient illustrates the strong reduction of the mutual interaction between cellular tips in the SOE regime due to the shift of stable $P_\lambda$.
This weak interaction between tips inhibits the organization of the cellular array inherited from the initial transient and is in line with the disorder that is observed experimentally and in simulations. Apart from the already mentioned dispersion of apparent cells' shapes in a transversal cut (Fig. \ref{transversal}), we would like also to mention the possibility, visible in Fig. \ref{3d}, for the cell's tip to be largely displaced with respect to the center of the cell. 
The large values of stable $P_\lambda$ also imply that the growth direction of the cells is prescribed by the anisotropy of interface energy (the drift velocity observed in the simulation in Fig. \ref{3d} is around 90\%  the one expected for a growth direction fully aligned with the cubic axis, i.e. in the minimum stiffness direction), and not by the direction of the thermal gradient. This supports the experimental observations in Figs. \ref{melt_pool} and \ref{melt_pool_zoom}. First, it explains why  the growth directions in regions 1 and 2 differ significantly (we have checked using EBSD that, indeed, the orientation of their crystalline axis differ). Second, as already mentioned, the presence at the center line of the melt pool of grains that have grown in a stable manner throughout the whole melt pool solidification process is rather unusual compared to common welding microstructures \cite{welding_book} and indicates a strong stabilization of dendrite tips due to anisotropy.
This question refers to grain competition and we  see that the particularities of the SOE regime of directional solidification may have important implications in this respect.
From a purely applicative point of view, the understanding of grain competition is even more important than the understanding of the intra-granular solidification microstructure. Indeed, heat treatments are usually designed so as to erase the microsegregation inherited from solidification within the grains. Typically, their duration $t$ is fixed according to the cell spacing $\lambda$ in order to allow for solid-state diffusion to completely homogenize the chemical composition, i.e. $t \propto \lambda^2$. 

\begin{figure}[htbp]
\includegraphics[angle=0,width=\linewidth]{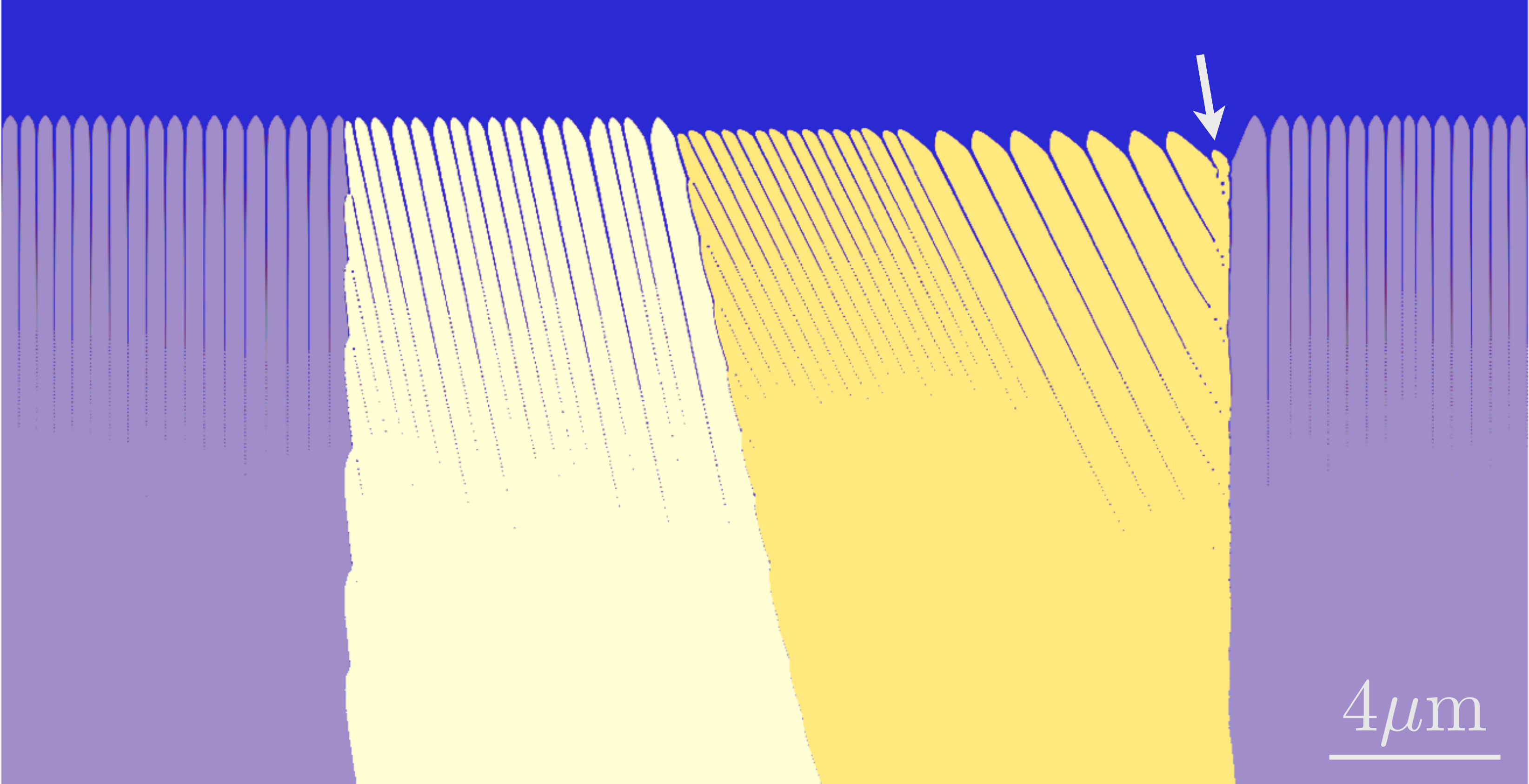}
\caption{\label{grains} Poly-crystalline solidification simulation with each color representing a different crystalline orientation. At the divergent grain boundary, the tertiary branching mechanism exhibited by the arrow produces new cells (on the right side of the yellow grain) with a spacing that is approximately three times larger than the one inherited from the transient (on the left side of the yellow grain).}
\end{figure}

Let us finally mention in this respect an interesting particularity of the SOE regime concerning the technologically relevant poly-crystalline solidification, for which grain selection for example takes place \cite{korner1, korner2}. In this case, tilted arrays of growing dendrites or cells coexist yielding grain-boundaries in the solidified material. At a divergent grain-boundary, the tertiary branching mechanism repeatedly takes place, and the spacing between the new cells is close to the maximum of the stable range. 
It was shown in Ref. \cite{karma_new} that, after a transient where the developed spacing is close to the minimum of the stable range, the new cells invade the grain, with an advection of the spacing distribution. The spacing between two given cells evolves in time due to the spatial variations of the drift velocity $V_d = V_g \tan \alpha$ ($\alpha$ represents the local growth direction), that itself depends on the spacing through $P_\lambda$ and is related to the anisotropy. 
When $P_\lambda \gg 1$, $V_d$ converges to $V_d^0 = V_g \tan \alpha_0$, the drifting velocity corresponding to a growth direction $\alpha_0$ aligned with the cubic axis of the crystal. 
When $dV_d/d\lambda$ is sufficiently large, the advection of the $\lambda$-distribution may take place at a speed significantly smaller than the drifting velocity \cite{karma_new}. In this regime the spacing between two given cells evolves with time and $\alpha$ is neither close to 0 nor close to $\alpha_0$, i.e. $1- \alpha/\alpha_0 \sim \alpha/\alpha_0 \lesssim 1$.
In the SOE regime, $\alpha/\alpha_0$ is close to 1 (as for the three-dimensional simulation in Fig. \ref{3d}) and $dV_d/d\lambda$ is small due to  large $P_\lambda$. The spatial variations of $V_d$ are small and the spacing between two given cells is constant in time. Then the drift velocity, close to $V_d^0$, and the speed of the advection coincide.
This regime is illustrated in Fig. \ref{grains} with a two-dimensional simulation of poly-crystalline solidification, with each color referring to a different crystalline orientation. Here, $V_g = 8$ cm/s, $G=2 \times 10^5$ K/cm and the grid spacing is 4 nm (otherwise the same parameters as in Fig. \ref{3d}).
The tertiary branching mechanism at a divergent grain boundary, exhibited with an arrow, produces a sequence of new cells (on the right side of the yellow grain) with a spacing as large as approximately three times the spacing of the old cells (on the left side of the yellow grain) inherited from the transient. Note that the possibility for this large difference should have an impact on the choice of the duration of a heat-treatment designed to erase the microsegregation. 
The large spacing invades the grain, and the transition between the small and the large spacing regions is very sharp, i.e. occurs on the scale of one cell spacing. 
According to the direction of the liquid channels, the growth direction does not change in time and is the same for the thin and the wide cells. Here, the spacing distribution is thus advected in a shape-preserving manner with the drifting velocity $V_d$ close to $V_d^0$.

\section{Conclusion}

As a conclusion, we propose the strongly out-of-equilibrium (SOE) regime of columnar directional solidification for which the dendrite P\'eclet numbers $P_\rho$ and the spacing P\'eclet number $P_\lambda$ are of order unity. 
Owing to $P_\rho/P_\lambda \sim 1$, side branching is strongly reduced. 
Moreover, in comparison to the weakly out-of-equilibrium (WOE) regime for which $P_\rho \ll 1$, the band of stable spacings is shifted to larger values of $P_\lambda$. This lowers drastically the ability of the array to organize via the phase diffusion process and yields a growth direction prescribed by the anisotropy of interface energy.  
SOE directional solidification thus takes place through  irregular arrays of unbranched dendrite/cells leaving an irregular microsegregation pattern in the solidified material, in accordance with our experimental observations of a as-built LPBF-processed IN718 alloy.

\section{Acknowledgements} 

We acknowledge funding by the German Federal Ministry of Education and Research in the framework of the "Forschungscampus Digital Photonic Production: DPP Direct", FKZ 13N13709.

\end{document}